\documentclass[acmlarge, nonacm]{acmart}

\AtBeginDocument{%
  }

\usepackage{xcolor}
\usepackage[most]{tcolorbox}
\usepackage{soul}
\usepackage{tabularx} 
\usepackage{colortbl}

\usepackage[normalem]{ulem}
\usepackage{multirow} 
\usepackage{array} 

\definecolor{highlight}{rgb}{1, 0.7686, 0}
\definecolor{myblu}{rgb}{0.0588, 0.4824, 1}
\definecolor{mypink}{rgb}{1, 0.4, 0.6}
\definecolor{deepblu}{rgb}{0.173, 0.106, 0.620}

\newcommand{\draft}[1]{\textcolor{gray}{(#1)}}

\newcommand{\edit}[1]{%
    \tcbox[colback=highlight!80, colframe=yellow!100, arc=1mm, boxrule=0pt, boxsep=2pt, left=2pt, right=2pt, top=0.5pt, bottom=0.5pt, on line]{\textsf{\small #1}}}
    
\newcommand{\strike}[1]{\sout{#1}}

\setcopyright{none}
\renewcommand\footnotetextcopyrightpermission[1]{}
\begin{document}

\title[Rewriting the Self]{Rewriting the Self: User Perceptions of AI-Mediated Cognitive Reframing}
\author{Yuewen Yang}
\authornote{Both authors contributed equally to this work.}
\affiliation{%
  \institution{Information Science, Cornell University}
  \city{New York}
  \state{NY}
  \country{USA}}
\email{yy2228@cornell.edu}

\author{Olzhas Yessenbayev}
\authornotemark[1]
\affiliation{%
  \institution{Information Science, Cornell Tech}
  \city{New York}
  \state{NY}
  \country{USA}}
\email{oy38@cornell.edu}

\author{Anna R. Van Meter}
\affiliation{%
  \institution{Child and Adolescent Psychiatry, NYU Grossman School of Medicine}
  \city{New York}
  \state{NY}
  \country{USA}}
\email{anna.vanmeter@nyulangone.org}

\author{Tanzeem Choudhury}
\affiliation{%
  \institution{Information Science, Cornell Tech}
  \city{New York}
  \state{NY}
  \country{USA}}
\email{tanzeem.choudhury@cornell.edu}


\author{Daniel A. Adler}
\affiliation{%
  \department{Computer Science and Engineering}
  \institution{University of Michigan}
  \city{Ann Arbor}
  \state{MI}
  \country{USA}}
\email{adlerdan@umich.edu}

\renewcommand{\shortauthors}{Yang et al.}

\begin{abstract}

Individuals are increasingly using AI for psychological support. One specific paradigm, AI-mediated cognitive reframing (AIM-CR), uses AI to automatically reframe users' negative thoughts (``\textit{I failed...}'') into more adaptive perspectives (``\textit{I am learning...}''). While prior work suggests that AIM-CR tools reduce users' negative emotions, the complexities of AI automatically modifying users' personal narratives remain underexplored. In this work, we investigated these complexities through a mixed-methods study in which 28 participants engaged with an AIM-CR tool that reframed narratives about their stressors. Our findings suggest that participants had mixed perceptions of AIM-CR: although reframing reduced negative emotions for most participants, participants felt less ownership of, and less authenticity in, the reframed narratives. Participants perceived AIM-CR as a validating ``mirror,'' as ``another self'' that could motivate or alienate them, or as an external social actor. These perceptions shaped therapeutic dynamics. We conclude by discussing design implications for future AIM-CR tools to better preserve user agency in narrative co-construction and more effectively support mental health and well-being.
\end{abstract}



\maketitle

\section{Introduction}

Artificial Intelligence (AI) is increasingly utilized to support psychological well-being, particularly by intervening in how individuals articulate and understand their personal experiences. One such application is AI-Mediated Cognitive Reframing (AIM-CR), where AI modifies a user's self-directed text (e.g., journal entries, reflections on stressors) to reflect alternative, more constructive viewpoints, based on principles from therapeutic Cognitive Reframing (CR) \cite{beck2011cbt, hollon2013cognitive}. For example, a self-critical statement like ``I'm a complete failure for missing that deadline,'' might be reframed as, ``It was a setback, but it's an opportunity to learn'' \cite{sharma_cognitive_2023, ziems2022inducing}. This approach builds on the established narrative-based therapeutic approaches that treat articulating and reframing experiences, especially challenging ones, as fundamental to meaning-making and psychological well-being \cite{pennebaker1997writing, Bruner1991-BRUTNC}.

The appeal of AIM-CR stems from its foundation in evidence-based CR practices \cite{beck2011cbt, hollon2013cognitive} and its practical advantages over more ``free-form'' therapeutic chatbots. While chatbots engage in open-ended dialogues that can get derailed over multiple turns \cite{10.1145/3715275.3732039}, AIM-CR offers a streamlined and focused experience by concentrating on the single, structured task of reframing negative thoughts \cite{sharma2024facilitating}. Initial evaluations have demonstrated promising benefits: AIM-CR systems can generate high-quality reframes \cite{ziems2022inducing, sharma_cognitive_2023, zhan2024large}, lead to reductions in negative emotions and their intensity \cite{sharma2024facilitating,fang2024leveraging, sharma_cognitive_2023}, and enhance resilience \cite{fang2024leveraging}, with some studies suggesting AI-generated reframes can even outperform human-written ones \cite{li2024skill}. 

However, despite promising technical and emotional benefits, there has been comparatively little investigation into users' perceptions of the \textit{AI mediation} itself. The act of AI modifying personal narratives introduces complexities beyond the immediate therapeutic outcome. Personal language expression serves crucial intrapersonal functions in self-regulation, emotional processing, and the construction of narrative identity \cite{vygotsky2012thought, Bruner1991-BRUTNC}. The integrity of these self-narratives is often linked to an individual's sense of authenticity—the congruence between internal experiences and expressed selves \cite{bauer2008narrative}. Unlike traditional therapy, where the therapist-client boundary is managed through a working alliance \cite{horvath1991relation}, AIM-CR directly reshapes the language expression of the user's emotions and experiences. Therefore, AI intervention in these deeply personal narratives, raises questions about the impact on user's identity, authenticity, and how this may affect therapeutic outcomes.

Within \textit{interpersonal} communication, the field of AI-Mediated Communication (AI-MC) has extensively studied how AI's textual interventions impact users \cite{hancock2020ai}. While AI can enhance efficiency and perceived communication quality \cite{coman2024perceptions, fu_text_2024}, it often introduces significant tensions. These tensions manifest as challenges to \textit{authenticity}, where users may be perceived as less sincere or find AI suggestions intrusive in their personal communications \cite{hohenstein2023artificial, fu_text_2024, liu2022will}. Additionally, there is a reduction in user \textit{agency} and \textit{ownership} over messages, making the text feel less like ``their own'' \cite{mieczkowski2022examining, buschek2021impact}, as users struggle to reconcile their personal expressive voice with AI's input \cite{fu_text_2024}.

It is not yet clear how these documented tensions translate to the unique \textit{intrapersonal, therapeutic} context of AIM-CR. How do users experience AI intervention in their personal narratives? Do users feel ownership over and commitment to the new reframed perspectives? How do users conceptualize AI's role in these interventions: does the resulting narrative still feel authentically their own, or does it feel like an external imposition? Investigating these questions is vital to the responsible design, translation and implementation of AI-assisted reframing interventions.

In the summer of 2024, we conducted a mixed-methods, single-session study with 28 participants using a technology probe. Participants wrote about a stressful experience and subsequently reviewed an AI-reframed version of their narrative generated by GPT-4o using CR principles. We collected quantitative data on emotional changes and perceptions of authenticity and ownership, alongside qualitative data from think-aloud protocols and semi-structured interviews. Our findings reveal a central tension: while AIM-CR often reduced negative emotions and increased alignment with participants' ``aspired selves,'' it simultaneously decreased feelings of authenticity to their ``current selves'' and their sense of ownership over the text. 
This tension was closely linked to how participants conceptualized the AI's role, forming distinct mental models of it as a validating ``Mirror,'' an aspirational yet intrusive ``Another Self,'' or a ``Social Actor'' that was scrutinized as an outsider's advice. These perceptions, influenced by the stylistic and emotional alignment of the reframes, influenced their therapeutic utility and acceptance.

This study makes several contributions. First, we provide empirical data on the emotional and psychological effects of AIM-CR, revealing a key tension between reducing negative emotions and maintaining authenticity to their narratives. Second, we explore AI's role beyond a simple automation of CR. We detail how users perceived various AI roles within the reframing process---the ``Mirror,'' ``Another Self,'' and ``Social Actor''---based on the perceived degree of alignment between the user and the AI-generated text, which shaped the therapeutic experience. Finally, based on this model, we offer concrete design implications for AIM-CR systems. We advocate for designing systems that can intentionally guide these mental models, empower users with co-creation and personalization controls to maintain agency and ownership, and build trust by incorporating user context. We argue that for AI to ethically and effectively support mental well-being, designers must account for the complex psychological dynamics of AI-mediated self-reflection.
\section{Related Work}

\subsection{Language and AI-Mediated Communication}
\label{related-language}

Language is fundamental to both \textit{interpersonal} communication and \textit{intrapersonal} self-understanding. \textit{Interpersonally}, language choices establish rapport by reflecting social contexts \cite{Brown_Levinson_1987}, while acts like self-disclosure build trust and common ground \cite{collins1994self, pickering2004toward, clark1991grounding, zwaan1998situation}. \textit{Intrapersonally}, language is the medium for regulating thoughts, understanding emotions, and interpreting experiences. People construct their self-identity by weaving life events into an evolving story; that story is continually evaluated for authenticity against their perceived ``true self'', revealing consistency or incongruence \cite{Bruner1991-BRUTNC, bauer2008narrative}.

Today, interpersonal communication increasingly integrates AI as an active mediator. Users employ AI-Mediated Communication (AI-MC) tools to generate and refine messages—making them more concise, professional, or friendly \cite{hancock2020ai,coman2024perceptions}. These tools have shown positive effects, increasing communication speed, positive affect, and cooperation \cite{hohenstein2023artificial}, and helping reduce anxiety about emotional expression and cultural nuance \cite{fu_text_2024}. However, despite these benefits, AI mediation introduces new complexities. Because language fulfills multiple roles---conveying information \cite{clark1991grounding}, expressing social dynamics \cite{Brown_Levinson_1987}, and constructing identity \cite{Bruner1991-BRUTNC}---integrating AI-generated language into communication has had a nuanced impact on users' emotions, agency and authenticity:

\paragraph{Emotional cost} AI can produce highly empathic messages, tempting users to outsource difficult emotional labor (e.g., apologies, condolences) \cite{yin2024heard}. However, recipients of such AI-generated apologies may find them hollow, demanding more personal cues before they can forgive \cite{lim2025effects, glikson2023ai}. Conversely, humanizing AI may increase perceived empathy but also fear of social judgment from it \cite{kim2022you}.

\paragraph{Concerns About Authenticity} Users often worry that AI-generated language misrepresents their personal ``voice''—the presence of the author's identity in a text \cite{tardy2012current}---creating an \textit{authenticity gap} where the text no longer reflects their identity or values \cite{hwang202580}. Users fear AI may flatten their individuality and report lower authenticity when writing informally to friends and family \cite{fu_text_2024}. Recipients of such AI-mediated messages also perceive reduced authenticity, especially for emotionally charged messages like apologies \cite{glikson2023ai}.

\paragraph{Reduced Ownership over Text} AI-assisted writing can diminish users’ authorship and psychological ownership over the text. Psychological ownership stems from self-investment, control, and familiarity \cite{pierce2003state}. These factors are weakened when AI generates content. For example, incorporating more AI suggestions reduced users' reported authorship over emails \cite{buschek2021impact}, and with fully AI-generated text, users reported no ownership at all \cite{draxler2024ai}. Users also felt less accountable for communication failures, blaming the AI instead \cite{hohenstein2020ai}.

\hfill \break 
 \hspace*{0.3cm} While existing research has primarily explored interpersonal (person-to-person) communications, AI is increasingly used in more private, intrapersonal contexts---specifically psychological interventions designed to shape internal dialogue and self-narratives \cite{sharma_cognitive_2023, sharma2024facilitating, fang2024leveraging, ziems2022inducing, zhan2024large, li2024skill}. This work aims to understand how users perceive AI mediation of personal narratives within these therapeutic contexts. In the following section, we review emerging research on this perspective, focusing specifically on AI-Mediated Cognitive Reframing (AIM-CR).

\subsection{Introducing AI Mediation into Therapeutic Reframing}
\label{related-AICR}

The words we use to describe our experiences shape how we \textit{feel} about them. Building on the profound connection between language expression and emotional well-being, therapeutic practices have long focused on helping people reframe their self-narratives in healthier ways \cite{beck_cognitive_1979, madigan2011narrative}. Cognitive Reframing (CR) is a clinical, structured technique focusing on turning maladaptive thoughts around failures or stressors, into more constructive perspectives centered on flexibility and growth \cite{clark2013cognitive}. This technique has proven clinically effective for reducing anxiety, managing stress, and improving motivation \cite{rothbaum2000cognitive, beck2011cbt}.

Recognizing the potential for increased accessibility and scalability, HCI researchers have explored using AI to facilitate such therapeutic language interventions. While some work focuses on psychotherapeutic chatbots \cite{heinz2025randomized, prochaska2021therapeutic, inkster2018empathy}, a growing area takes a more direct approach by integrating AI-powered Cognitive Reframing (CR) modules into digital mental health interventions. These systems automatically transform a user's negative thought (e.g., ``I failed the exam, I am a failure'') into a reframed version (e.g., ``I faced a setback, but it's a growth opportunity'') \cite{sharma2024facilitating}. In this paper, we refer to these tools as \textit{AI-Mediated Cognitive Reframing (AIM-CR)} systems \cite{sharma_cognitive_2023, sharma2024facilitating, fang2024leveraging, ziems2022inducing, zhan2024large, li2024skill}.

Importantly, AIM-CR is motivated by both therapeutic and practical considerations. AIM-CR provides an evidence-based therapeutic intervention in a streamlined form \cite{sharma2024facilitating}. This direct reframing avoids multi-turn dialogues with chatbots, which may not be as good at constraining user inputs and conversation flow, and may be derailed by the users into inappropriate conversations \cite{10.1145/3715275.3732039}. The underlying reframing logic can be based on various frameworks, including CBT-based cognitive distortion identification \cite{sharma_cognitive_2023, sharma2024facilitating} and positive psychology fostering gratitude and growth mindsets \cite{fang2024leveraging, ziems2022inducing}. AIM-CR tools have been recently deployed in mainstream mental health platforms reaching hundreds of thousands of users \cite{mha_2024a}.


The therapeutic outcomes from AIM-CR are promising. Initial evaluations with both users \cite{ziems2022inducing, sharma_cognitive_2023} and therapists \cite{zhan2024large} demonstrated that AI can produce high-quality, empathetic reframes. Recent studies have shown reduced negative emotions\cite{sharma2024facilitating} and increased commitment and resilience \cite{fang2024leveraging}. Some research suggests that LLM-generated reframes outperform human-written ones in reducing negative emotions, even when participants know the reframes are AI-generated \cite{li2024skill}. Researchers have also begun exploring novel delivery modalities, such as presenting audio reframes in users' own synthesized voices to enhance internalization \cite{fang2024leveraging}.

However, these evaluations have focused primarily on technical feasibility and psychological outcomes. They have given relatively little attention to users' perceptions of the \textit{AI mediation} itself. This gap is significant given what we know from AI-MC research, where AI mediation of user communications consistently creates complex effects that directly impact users' authenticity, ownership and identity (see Section \ref{related-language}). In AIM-CR, we may encounter similar complexities as AI modifies users' intrapersonal narratives. In this work, we focus on these perceptions within the AIM-CR, which we expand on in the next section.



\subsection{Identity, Authenticity, and Ownership Tensions in AIM-CR}
\label{related-AI-MC}

It is not yet clear how tensions around identity, authenticity and authorship, identified with AI interventions in interpersonal communications (Section \ref{related-language}), may manifest in or affect the therapeutic AIM-CR contexts. AIM-CR tools modify how users express their feelings towards and relate to life experiences. While the stated goal is to align the narratives with more adaptive ways of thinking, how do users feel about AI modifying their thoughts and emotional expressions? How does it affect their psychological ownership of the text (a feeling that they wrote it), and emotional authenticity---an important component of successful therapeutic outcomes\cite{mengers2014benefits}? In traditional therapy, these dynamics are managed through \textit{working alliance}—the collaborative bond between client and therapist\cite{horvath1991relation}---built on the client's trust, collaborative agency, and ownership over their therapeutic process---that is a key predictor of positive outcomes . These dynamics have not yet been explored with AI-mediated reframing. Introducing AI into users' internal sense-making, especially without robust safeguards, can lead to risks of delusional thinking \cite{10.1145/3715275.3732039}, and in the real world even prompted conspiratorial beliefs and suicidal ideation \cite{hill_chatbots, hill_they_asked}. These concerns create unique lines of inquiry around user perceptions of AI in AI-Mediated Cognitive Reframing interventions, specifically:

\paragraph{How do users experience AI intervention in their personal narratives?} Changes to users' self-narratives in AIM-CR are not merely stylistic changes as in most AI-MC studies (see Section \ref{related-language}), but semantic and affective transformations of narratives about personal stressors and life experiences \cite{sharma2024facilitating, fang2024leveraging}---content that is central to how individuals construct their sense of self \cite{Bruner1991-BRUTNC}. The scope of intervention therefore shifts from modifying self-presentation to potentially reshaping self-understanding. How do users feel about AI mediating not just their outward expressions, but their internal sense-making processes?

\paragraph{How does AI-mediated reframing affect feelings of authenticity?} Within AIM-CR, users have relatively limited control over AI. The prompts guiding the reframing are based on predetermined therapeutic frameworks, which may not be transparent to the end users \cite{sharma2024facilitating, ziems2022inducing}. How do users perceive the authenticity of AI reframes if they do not align with users' emotions?

\paragraph{How do users conceptualize AI's role in reframing their self-expressions?} Unlike traditional therapy where clients retain clear ownership over their narratives, AIM-CR creates ambiguity by reframing a user's own text. This raises questions about whether users still identify with the AI-altered narratives, which is critical given that psychological ownership over one's health journey is a determinant of long-term outcomes \cite{karnilowicz2011identity}. How users conceptualize the AI—as a therapist, an extension of the self, or an automated tool—will likely affect their therapeutic self-disclosure, acceptance, and internalization of the reframes \cite{sannon2018personification}.

\hfill 
\break 
 \hspace*{0.3cm}
Understanding users' perceptions of these dynamics represents a novel territory, which is crucial to investigate towards the effective design, translation, and implementation of novel AI psychotherapeutic interventions. To explore these questions, we conducted a mixed-methods study to understand perceptions of AIM-CR and inform its responsible design. The next section details our methods.

\section{Methods}

For our study, we developed a web application as our technology probe, which was used in the mixed-methods, single-session study to collect data and gain insights into user needs and experiences \cite{hutchinson_technology_2003, huang_technology_2014}. The study procedure is shown in Figure \ref{fig:study-procedure}. 

\subsection{Participants}
We conducted a single-session study with 28 participants. Participants were primarily graduate students (27 out of 28) recruited from universities in the northeastern United States via mailing lists and university-affiliated online forums.
The sample was predominantly young adults: 50\% (n=14) were in the 18-25 age group, and the other 50\% (n=14) were in the 26-34 age group. 14 participants identified as male (50\%), 12 as female (42.9\%), and 2 as non-binary (7.1\%).
No participants reported current diagnoses or treatment for mental health conditions, such as depression and anxiety. All participants provided informed consent before the study. This study was approved by the authors’ Institutional Review Board (IRB). Each participant received a ten-dollar Amazon gift card as compensation for their time (approximately 1 hour).

\subsection{Protocol}

We employed a mixed-methods approach, collecting quantitative data through surveys and qualitative data through think-aloud sessions and semi-structured interviews. 

\subsubsection{Measures} We assessed (1) in-the-moment emotion and the corresponding intensity, (2) positive and negative affect (PANAS) \cite{diaz-garcia_positive_2020}, and (3) authenticity and ownership of the written text. The emotional measures were selected based on previous research on single-session interventions for mental health \cite{schleider_harnessing_2022}, as well as self-guided AI-based interventions \cite{sharma2024facilitating}. There is evidence that even brief online interventions can yield significant emotional benefits \cite{SCHLEIDER2017107}. The authenticity and ownership measures were informed by prior work highlighting these concepts as crucial for understanding the therapeutic effects of AIM-CR tools (See Section \ref{related-AI-MC}). Specific questions can be found in Appendix \ref{appendix-measures}.

\subsubsection{Study Procedure} 
\begin{figure}
    \centering
    \includegraphics[width=1\linewidth]{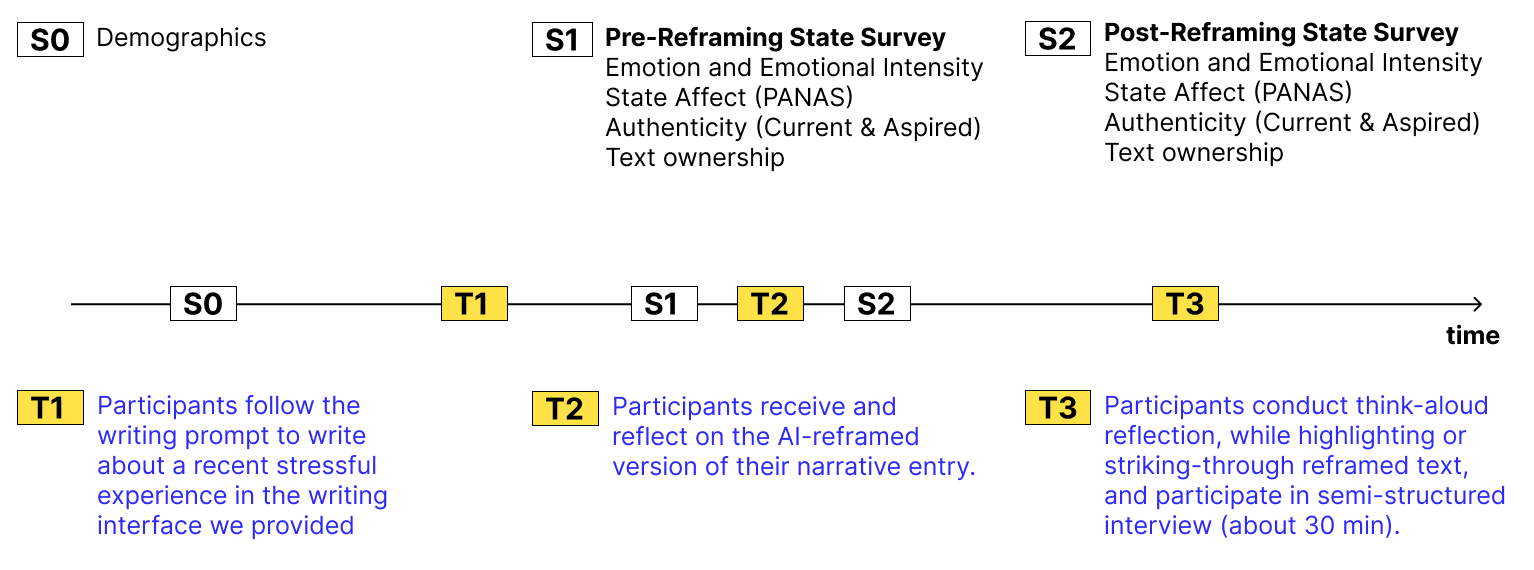}
    \caption{Overview of the study procedure, illustrating the sequence of participant tasks and the corresponding survey administrations (S0, S1, S2) with measures collected at each time point. Our partcipants' tasks include initial writing (T1 in figure), reviewing the AI-reframed narrative (T2 in figure), and the think-aloud reflection/interview (T3 in figure).}
    \label{fig:study-procedure}
\end{figure}

Each session involved several key tasks. First, participants completed baseline questionnaires, including demographic information (Fig. \ref{fig:study-procedure} S0). They then authored a personal narrative about a recent stressful experience using the web-based writing interface (Fig. \ref{fig:study-procedure} T1). The specific writing prompt could be found in Appendix \ref{appendix:writing-prompt}, which was adapated from Pennebaker's Expressive Writing prompt \cite{pennebaker_expressive_2014}. After completing a post-writing survey on their current emotional state and perceptions of their own text (Fig. \ref{fig:study-procedure} S1), participants were presented with an AI-reframed version of their narrative. After receiving and reflecting on the AI-reframing, participants filled out the post-reframing state survey based on their feelings (Fig. \ref{fig:study-procedure} T2, S2). 
Participants were then asked to verbalize their thoughts and reactions while using interface tools to highlight, strike-through, or edit portions of the AI's output. This allowed us to understand participants' reactions to and perceptions of AI-reframed self-narratives in a fine-grained way. Lastly, we conducted semi-structured interviews to get in-depth qualitative feedback on their overall experience (Fig. \ref{fig:study-procedure} T3).

Specific interview questions were organized around three primary aspects, and follow-up questions were asked for adaptation and probing:
    (1)``\textit{How did you feel about the AI's modifications to your personal narrative?}''
    (2)``\textit{Walk us through your rationale of each highlighted, striked-through, or edited text}''; and (3)``\textit{How did you navigate between the AI's perspective and your own when reviewing the modified text?}''


\begin{figure}[h]
    \centering
    \includegraphics[width=1\textwidth]{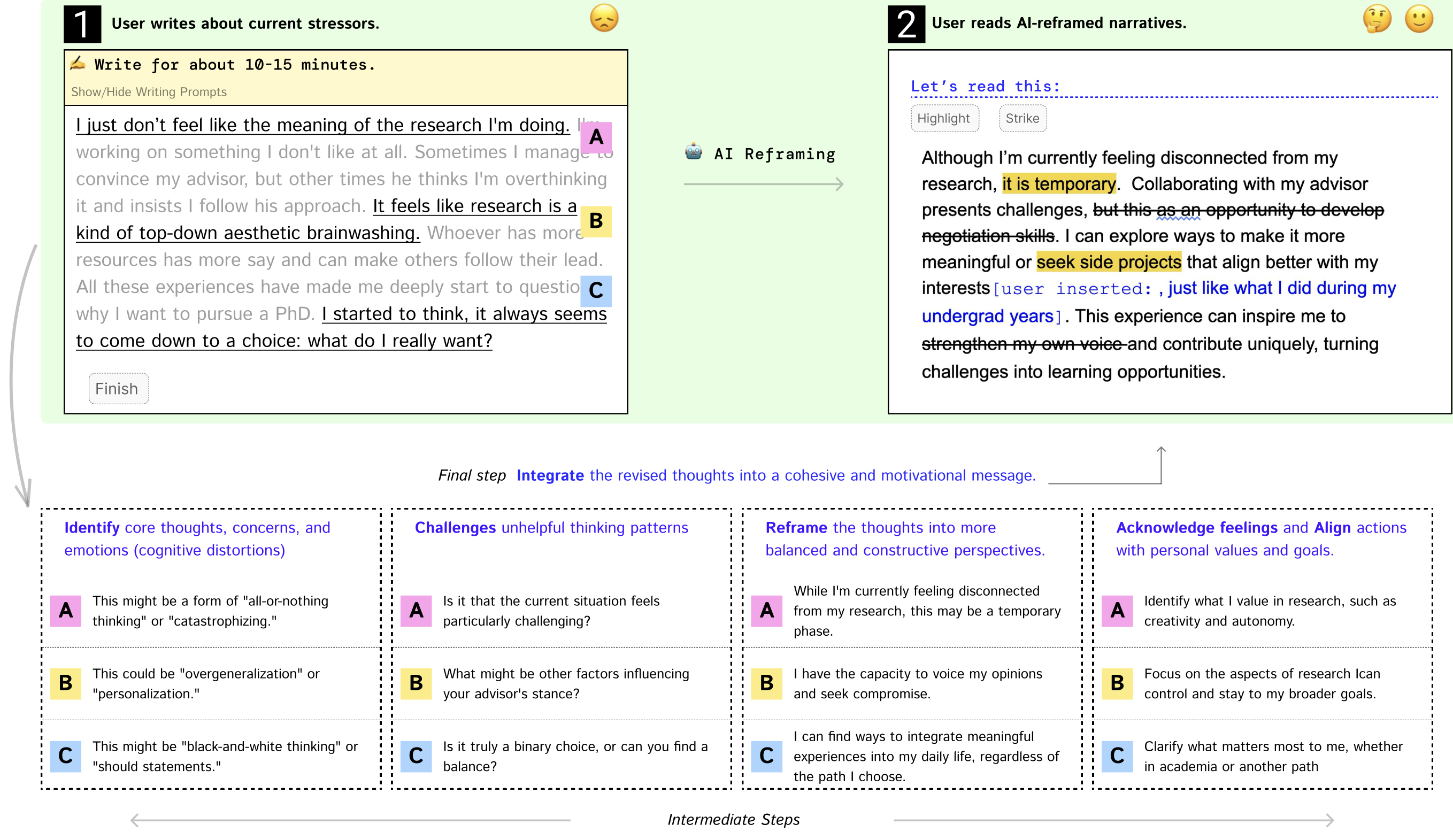}
    \caption{\textit{\textbf{ Interface and Workflow of the Technology Probe for AI-Mediated Cognitive Reframing.}}
This figure demonstrates the user interface and AI reframing workflow of the technology probe. The upper part shows the technology probe interface, featuring (1) a writing textbox for user input and (2) a reframed narrative display for interaction (highlight, strike-through, edit) during think-aloud sessions. The user-entered text, shown in the box on the upper left, is processed through a series of prompts in the GPT-4o model to generate reframed narratives. The lower part illustrates the AI reframing process, detailing hidden intermediate steps (e.g., identifying cognitive distortions, challenging patterns, reframing thoughts) that generate the final narrative, which is then displayed in the interface. The intermediate steps are not visible to the user.
}
    \Description{This figure illustrates the workflow of our technology probe (TP).}
    \label{fig:reframing-flow-and-interface}
\end{figure}

\subsection{ Writing Interface for AI-reframing}
\label{subsection:writing-interface}
\subsubsection{Prompting AI for Narrative Reframing}

To reframe user narratives, we used Cognitive Reframing (CR), a well-established psychological technique \cite{beck_cognitive_1979} previously explored in research on LLM-assisted reframing \cite{sharma_cognitive_2023, sharma2024facilitating}. We prompted a GPT-4o model to follow the CR process: \textbf{(1)} identifying core thoughts, concerns, emotions, and cognitive distortions; \textbf{(2)} challenging maladaptive thinking patterns; \textbf{(3)} reframing thoughts into more balanced and constructive perspectives; and \textbf{(4)} acknowledging feelings and aligning actions with personal values and goals. This process was implemented using a chain-of-thought prompting strategy \cite{wei_chain--thought_2023}, breaking the overall task down into smaller, sequential prompts. Few-shot examples of CR were provided for each step to guide the model's output \cite{brown_language_2020}. The model was also prompted to maintain the user's original writing style. Full details are available in Appendix \ref{appendix:ai-prompt}. The user was only presented with the final rewritten narrative. Fig. \ref{fig:reframing-flow-and-interface} illustrates the workflow of our technology probe.

To refine the reframing process and gather preliminary feedback, we conducted a pilot study with five participants (not included in the full study with the 28 participants). This pilot study indicated that the system could produce coherent reframed text that participants found engaging.

\subsubsection{Technology Probe Interface}
The system provided a rich-text editing interface that enabled users to (1) highlight helpful reframes, (2) strike through unhelpful ones, and (3) edit the reframed text during \textit{think-aloud sessions} (Fig. \ref{fig:reframing-flow-and-interface}). These actions were logged for later analysis. The system was deployed as a web application, allowing remote participants to access it.

\subsection{Data Collection and Analysis}

Each session was conducted via Zoom and recorded with participants’ permission. Participants were explicitly instructed not to include highly sensitive personal information (e.g., full names, specific medical diagnoses, or identifiable details) in their written narratives, to reduce the risk of unintended disclosure. Survey data was collected via Qualtrics during the session.

A mixed-methods approach was employed to analyze the collected data. For quantitative analysis, to assess the impact of CR, we calculated descriptive statistics and scores assessing the change in measures pre- versus post-CR (S1 versus S2 in Fig.\ref{appendix-measures}). To analyze changes in emotional states and narrative perceptions following the AI-mediated cognitive reframing, we calculated delta scores (post-reframing minus pre-reframing) for each measure. 
We presented descriptive statistics including mean delta scores and the percentage of participants showing improvement or no change for each measure, as visualized in Fig. \ref{fig:emotional} and Fig. \ref{fig:authenticity}. Specifically, PANAS positive and negative affect scores were calculated as the mean of five items each (positive: Active, Determined, Attentive, Inspired, Alert; negative: Afraid, Nervous, Upset, Hostile, Ashamed). Meaningful change thresholds were defined as 0.5 standard deviations of the pre-reframing scores \cite{norman_interpretation_2003}. Participants were categorized as showing meaningful positive change (above $+0.5$ SD), meaningful negative change (below $-0.5$ SD), or negligible change (within $±0.5$ SD).

Audio recordings from the think-aloud protocols and semi-structured interviews were transcribed via Zoom recording transcriptions provided by the authors' institution. The first two authors analyzed the transcriptions qualitatively using a reflexive thematic analysis approach adopted from \cite{braun_using_2006}, combining both inductive and deductive elements \cite{adler_beyond_2024, braun_one_2021}: codes and themes arose from the data, but were guided by our inquiries raised and literature synthesized in the related work. The two researchers independently coded a subset of the data, with discrepancies resolved through discussion to finalize the coding themes and coded the rest. The quantitative and qualitative findings were then integrated to provide a comprehensive and nuanced understanding of the user experience.

\subsection{Positionality Statement}

At the time of conducting the study and working on this project, the two first co-authors are graduate students in computing and information science at a university in the Northeastern United States.
These authors conceptualized and planned the study, analyzed collected study data, and were primarily responsible for writing this paper.
One author is a clinical researcher and practicing mental health clinician with expertise in cognitive reframing interventions.
This author provided feedback and revisions on the paper from a digital mental health perspective. 
One author is a faculty member of computing and information science at the same university as the two first co-authors, who provided supervisory support.
The final author is a senior computing and information science graduate student, who supervised the two first co-authors, and provided feedback on and revisions.
\section{Findings}

This section presents participants' experiences with AI-mediated cognitive reframing (AIM-CR) of their personal stressors. Through a primarily qualitative lens, supported by descriptive quantitative data (Figures \ref{fig:emotional} and \ref{fig:authenticity}), we unpack our findings in three parts. First, we explore the therapeutic effects of AIM-CR, detailing how it decreased negative emotions and offered new perspectives, and examine the qualities that made reframes successful. Second, we investigate the complex relationship between participants' authenticity and ownership over the text and AI-generated first-person narratives, revealing a tension between their ``current-self'' and ``aspired-self'' authenticity. Finally, we analyze how participants conceptualized the AI's role within the reframing process—as a ``mirror,'' ``another self,'' or a ``social actor''—and how these mental models influenced their engagement with the reframed texts.

\subsection{Effects of Cognitive Reframing}
\label{subsection:effects-of-cr}

\subsubsection{Decreased Negative Emotions and Emotional Intensity}

\begin{figure}
    \centering
    \includegraphics[width=1\linewidth]{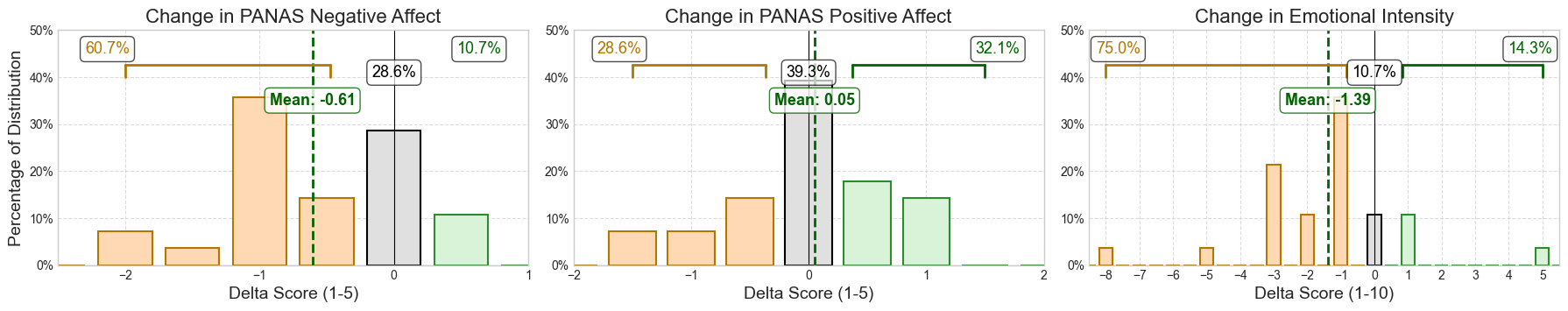}
    \caption{\textbf{\textit{Changes in Emotional Measures.}} This figure summarizes how participants' emotional states shifted following cognitive reframing (CR). Each panel presents the percentage distribution of participants (n=28) across different change scores, calculated as post-CR minus pre-CR values. The left panel shows changes in PANAS negative affect (1-5 scale), with a mean change of $-0.61$, reflecting an overall reduction in negative emotions. The middle panel displays changes in PANAS positive affect (1-5 scale), and the right panel illustrates changes in emotional intensity (1-10 scale). Green bars indicate positive changes ($\Delta>0$), orange bars indicate negative changes ($\Delta<0$), and gray bars represent minimal or no change (for PANAS, within $±0.5$ SD of pre-CR scores). Dashed vertical lines mark the mean change in each panel, and colored brackets highlight the range of positive and negative changes. Percentage labels denote the proportion of participants in each change category.}
    \label{fig:emotional}
\end{figure}

We analyzed pre- and post-intervention changes in negative and positive affect (PANAS) and emotional intensity (Figure \ref{fig:emotional}). CR led to reductions in negative emotions and overall emotional intensity. Specifically, for PANAS negative affect (1-5 scale), we observed a mean change of -0.61 (decrease reported by 60.7\% of participants). Emotional intensity (1-10 scale) also decreased, with a mean change of -1.39 (decrease reported by 75.0\% of participants). In contrast, changes in PANAS positive affect (1-5 scale) were smaller. While there was a slight average increase (Mean = 0.05), only 32.1\% of participants reported an increase, whereas 28.6\% reported a decrease and 39.3\% reported no change.

\subsubsection{Introducing New Perspectives}

The central goal of cognitive reframing is to help individuals see their situations from a new, adaptive perspective. Our qualitative data illustrate how AI-generated reframes accomplished this by offering alternative interpretations. P16’s experience exemplifies this process, who appreciated AI's interpretation:

\begin{quote}
    \textit{``I've always believed that every situation can be interpreted in many different ways. When the AI framed my situation, I thought---yeah, that's one valid interpretation. It reminded me that even something I experienced firsthand can be seen from another angle.''} (P16)
\end{quote}

This helped P16 see her situation as a \textit{``chance to do something differently,''} which \textit{``helped shift my mindset toward something more optimistic''} and \textit{``made me think ... this could be an opportunity to handle the situation better.''} For P1, the new perspective was about helping him \textit{``extend my own thinking''} by predicting \textit{``some things that are correct and make sense in context, even though I hadn't considered them.''} The reframes also served as reminders of things they may have missed: P18 was reminded that there are \textit{``a lot more ways I can communicate and stay connected with my parents,''} and P21 found it helpful that the AI is \textit{``trying to remind me to be more empathetic to how someone else is experiencing something.''}

\subsubsection{Positivity, Encouragement and Self-Compassion}

The reframes often embedded constructive suggestions or problem-solving tips, which participants found encouraging and motivating. For instance, P24, who was wrestling with self-blame, found that reading an AI-reframed story didn't make the stress vanish, but it did lead to a softening of her self-criticism and offered reassurance:

\begin{quote}
``\textit{[AI-reframing] gave me some reassurance---you already did something, and not achieving the goal is due to certain reasons, or how you can do better next time.}''
\end{quote}

For some participants, the AI fostered compassion by helping them reinterpret their struggles and recognize their efforts. P22, for example, described his failure to help his neighbor. He noted that a reframe allowed him to see this not solely as a personal shortcoming, but as part of a shared journey with the entire community: 

\begin{quote}
\texttt{\textbf{Reframe}  \hspace{1em} Together with my community, we can act in ways that reflect our shared values of compassion, justice, courage, and unity, striving towards a safer environment for everyone.} \\

\textbf{P22}: \hspace{1em} \textit{``It reinforces the idea that the actions I took weren't all for nothing, that they did have an impact. It also makes me feel better because it makes me feel less alone in the situation, identify with the neighbors as allies as opposed to just a bunch of random people.''}
\end{quote}

In moments of high distress, some participants craved even more direct expressions of comfort. P12, reflecting on what would have helped him, wished for the AI to offer simple, compassionate words: \textit{``Saying something like, `You've already done very well; you've tried your best.' That would comfort me more.''}

\subsubsection{Reframe Qualities}

The perceived utility of reframes often depended on their actionability, specificity, and controllability. Reframes that provided a clear sense of direction were highly appreciated, such as those that felt like a ``\textit{solution-oriented text}'' (P23) or effectively turned emotions into ``\textit{to-dos}'' (P21). This effect was strongest when suggestions were grounded in tangible examples. For instance, for P5, a graduate student struggling to find enjoyment in his research and internship search, a reframe suggesting ``\textit{exploring side projects}'' resonated deeply, as it reminded him of positive past experiences: \textit{``When I was an undergrad I did this a lot. I was way happier.''} To this end, he wished for even more personalization with relevant examples, suggesting: \textit{``I wish AI could provide more elaboration by offering examples, such as `I remember a student I know who managed to travel around the world while doing their research'. ''}

However, even when suggestions were relevant, participants navigated a fine line between helpful specificity and unhelpful rigidity. While concrete language made reframes easier to accept (P15: ``\textit{If it has concrete actions, then yeah, that's a good idea}''), overly directive commands were rejected. P16 emphasized the importance of choice, suggesting that having \textit{``multiple options''} would be more effective than a single directive: ``\textit{If the AI says `You have to do this'---as a [one fixed] solution---I'd definitely not go for it. If the AI gave me more options, I would be more receptive to that}.''

Finally, the value of actionable reframes diminished when they were incompatible with personal constraints or situational realities, and caused frustration. For example, P12 wrote about stressors related to his visa status, which prevented him from visiting his family. The AI's reframe suggested possibilities---\texttt{``I can find ways to stay connected and supported,''}---but because the core issue was outside his control, he dismissed this as invalidating ``\textit{empty talk}.'' This highlighted a need for the AI to have sufficient context; otherwise, its advice risked feeling irrelevant.

\subsection{Navigating Identity and Authenticity in AI-Reframed Narratives}

A key aspect of AIM-CR is that reframes are integrated within the user's first-person narratives. It embeds new perspectives directly into users' own self-expressions rather than through a multi-turn conversation. In our study, we observed that this approach had complex effects on how participants engaged with their reframed narratives, particularly in relation to authenticity and identity. Participants consistently evaluated the AI-generated texts against their own situations, emotional truth, and self-concept, with their acceptance of the reframes hinging on how ``authentic'' the text felt—both in terms of language and emotional resonance.

\subsubsection{Effects on Authenticity}

\begin{figure}
    \centering
    \includegraphics[width=1\linewidth]{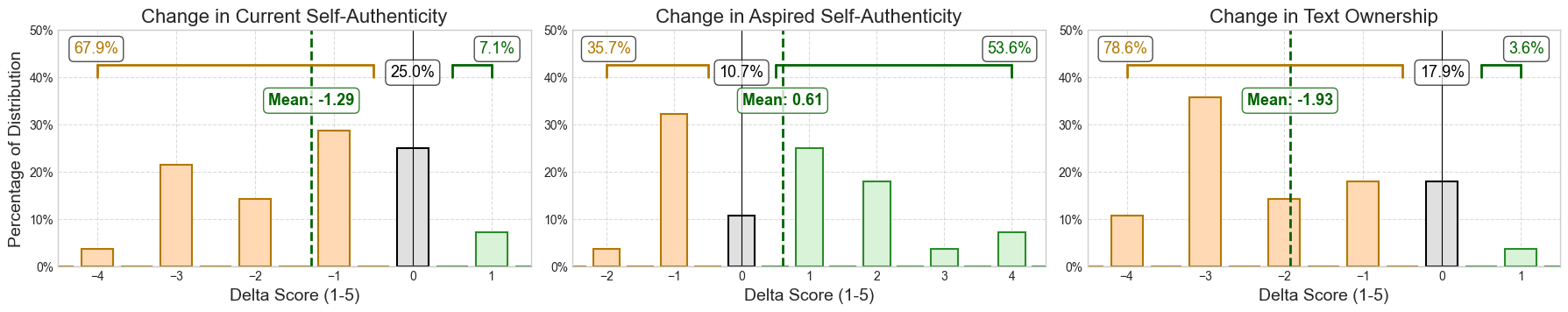}
    \caption{\textbf{\textit{Changes in Authenticity and Text Ownership.}} This figure summarizes how participants' perceptions of authenticity and text ownership shifted after receiving AI-reframed narratives. Each panel presents the percentage distribution of participants (n=28) across different change scores, calculated as post-intervention minus pre-intervention values. The left panel shows changes in current self-authenticity (1-5 scale), with a mean change of -1.29, reflecting that participants felt the reframed text was less aligned with their current self. The middle panel displays changes in aspired self-authenticity (1-5 scale), and the right panel illustrates changes in text ownership (1-5 scale). Green bars indicate positive changes ($\Delta>0$), orange bars indicate negative changes ($\Delta<0$), and gray bars represent no change ($\Delta=0$). Dashed vertical lines mark the mean change in each panel, and colored brackets highlight the range of positive and negative changes. Percentage labels denote the proportion of participants in each change category.}
    \label{fig:authenticity}
\end{figure}

We found complex effects on authenticity and text ownership post-CR (see Figure \ref{fig:authenticity}). While reframed narratives aligned more closely with participants' ideal selves, they felt less representative of their current feelings and sense of ownership. Specifically, for \textit{aspired self-authenticity} (1-5 scale), we observed a mean increase of +0.61, reported by 53.6\% of participants. In contrast, \textit{current self-authenticity} (1-5 scale) decreased, with a mean change of -1.29, reported by 67.9\% of participants. Similarly, psychological \textit{ownership} over the text (1-5 scale) also saw a substantial reduction (Mean = -1.93), with 78.6\% of participants reporting a decrease. This suggests a key tension: while reframes could feel aspirational, they often created a disconnect from participants' immediate sense of self.

\subsubsection{Increased Aspired-Self Authenticity}
\label{finding-authenticity-aspired}

The resonance between the reframed text and the perceived ideal self could be empowering, fostering a sense of progress towards a desired identity. P23, for example, found validation in this alignment, stating, \textit{``...seeing it's reflected here, just makes me feel that I have embraced the thought process that is gonna help me be the person that I would like to be.''} Similarly, P3 found that the reframed narrative presented aspired qualities, outlining a path for professional and personal development. While acknowledging that the AI might sometimes exaggerate benefits, P3 recognized a projection of their potential future, one they could grow into:

\begin{quote}
    \textit{``[The AI] reframes it in a way that highlights my skills and personal growth. It presents it as a chance for me to grow into those bigger roles... I do believe that even these specific problems can contribute to high-level benefits, like leadership development. That's a big commitment, and I'd say... yeah, I can see myself growing in that direction.''} (P3)
\end{quote}

This empowering effect was also evident when the AI embedded constructive suggestions. For example, P23 noted,\textit{``When I read it out loud, it sounds like me affirming those things—like, I will do this''}, which propelled them to take action to resolve the situation. Similarly, P6 said \textit{``It feels like I'm the one writing it, giving myself guidance''}.

\subsubsection{Reduced Current-Self Authenticity}
\label{finding-authenticity-current}

On the other hand, participants felt that the reframed text was less representative of their \textit{current self}, and reported reduced \textit{current-self authenticity} and psychological \textit{ownership} over the text (Figure \ref{fig:authenticity}).

\paragraph{Adapting AI's Suggestions}

After filling the post-CR authenticity questionnaires, participants could use the editing features of our design probe (see Section \ref{subsection:writing-interface}) to better align the text with their own preferences. Their edits reveal the specific sources of misalignment that could lead to the decrease in current-self authenticity. Participants made targeted changes---such as softening assertive or overly positive language---to reclaim their narrative and align it with their present emotional reality. These adaptations, illustrated in Figure \ref{fig:participant-edits}, highlight the friction between the AI's suggestions and the participants' felt experience, which we analyze in the remainder of this section.

\begin{figure}[ht]
\centering
\begin{tcolorbox}[colback=gray!5!white, colframe=gray!60!black, title=Participant Edits to AI-Generated Reframes]
\begin{quote}
\texttt{\textbf{P8} \hspace{1em} Acknowledging my discomfort without judgement, I \strike{commit} $\rightarrow$ \edit{try to} to embrace my unique identity and values.} \\
\hspace{1em} (\textit{``Committing is hard... That's not something that I feel confident to say out loud...It's just this is so certain that ... I don't feel confident saying that.''})  \\[1em]

\texttt{\textbf{P10} \hspace{1em} We ended up spending five \strike{enjoyable} hours together and connected on various topics. Despite feeling a bit awkward at first, \strike{I did my best} to stay relaxed as we caught up.} \\
\hspace{1em} (\textit{``The AI model interpreted some stuff that I wrote wrongly.''}) \\[1em]

\texttt{\textbf{P15} \hspace{1em} Although I have felt detached from the academic community and experienced FOMO, \strike{I can re-engage through online forums and local meetups}.} \\
\hspace{1em} (\textit{``The tone is unnatural to me. This is not a way in which I would write things.''}) \\[1em]

\texttt{\textbf{P20} \hspace{1em} Although I sometimes doubt the quality of my output, I can \strike{break tasks down into smaller, manageable goals}.} \\
\hspace{1em} (\textit{``It's the kind of the advice you always do.''})
\end{quote}
\end{tcolorbox}
\caption{Examples of participants editing AI-generated reframes to better align with their authentic voice and emotional state. Strikethrough indicates \strike{deleted} AI text; highlight indicates participants' \edit{insertions}. Participants' \textit{``quotes''} state their reasoning for these edits.}
\label{fig:participant-edits}
\end{figure}

\paragraph{Personal style}  A key reason participants sometimes felt disconnected from the reframed narratives was a mismatch in personal language and stylistic choices. Participants were sensitive to how the AI expressed it: it often lacked the nuance or emotional cadence they would use themselves. For example, the AI's language felt ``\textit{robotic}'' and ``\textit{too grand}'' (P4), ``\textit{way too healthy}'' (P8), resembling slogans that lost the rawness of their original feelings. Some found the tone ``\textit{too optimistic or fluffy}'' (P25). This stylistic misalignment could undermine the sense of authenticity, even when participants recognized the reframed text as helpful or valid. As P15 explained: \textit{``[The reframes are] understandable and valid, but it's not a way in which I process. The tone is unnatural to me. This is not a way in which I would write things.''}

\paragraph{Emotional (mis)alignment}

Beyond stylistic mismatches, participants experienced deeper emotional misalignment when the AI simplified their narratives, clashing with their desire to preserve emotional complexity. Many saw their writing as a space for honesty and processing, so mismatches could disrupt word choice, the overall structure (P27) and metaphor usage (P17). For those who used writing to capture nuanced emotional journeys, the AI’s tendency to smooth over or resolve tension felt especially jarring. As P17 described, the AI’s reframe erased the ``jagged parts'' of her story that were essential to its meaning:

\begin{quote}
    \textit{``I tend to use a lot of metaphor, and I go more in depth into the details---especially when I'm talking about something like stress. [...] I was describing all the little things I was feeling in each moment... so many nuances. But when the AI reframed it, it kind of compressed all of that. It simplified it too much. It felt like it got reduced to just, `Oh, it had a happy ending.' What I wanted was to capture the tension---the worry I carried right up to the very last moment, and all the struggles and roadblocks I faced along the way. The rewrite... smoothed over all the jagged parts I wanted to keep.''} (P17)
\end{quote}
This desire to preserve the authentic emotional arc of a story was echoed by P24, who explained that her expectation for narratives is to ``\textit{retain my original emotions},'' making her own writing distinct from the AI's.

\paragraph{Unrealistic expectations from the aspired self} On the other hand, the perceived gap between the aspired-self portrayed by the AI and the user's current-self view could increase stress rather than alleviate it. For example, P12 noted that \textit{``It gives me unrealistic expectations or makes me feel inadequate. It expects me to have steps I can take, but I really can't.''}. Similarly, P4 remarked excessive positivity was counterproductive:

\begin{quote}
    \textit{``It's not that I don't know [how to do it] --- but more that I don't want to do it, or I can't do it. It feels too positive, making my dark side seem even darker.''} (P4)
\end{quote}

\subsection{Conceptualizing the AI's Role in Cognitive Reframing}
\label{findings-ai-voice}
\label{finding-ai}

In this section, we unpack how participants made sense of the AI's role in the reframing process. While AI-mediated reframing could be viewed as a simple automation of a therapeutic technique, the reframed perspectives were presented in the participant's own voice but contained new, AI-driven perspectives. This prompted participants to form distinct mental models to interpret the ``voice'' and agency behind the reframed text. For participants, AI functioned as (1) a mirror, automating cognitive processes to reflect and validate their own thoughts; (2) as an aspired or ``another'' self; or (3) as a distinct social actor subject to interpersonal norms. These conceptualizations shaped the therapeutic experience beyond CR effects (see Section \ref{subsection:effects-of-cr}), influencing whether reframes were internalized as self-generated insight or received as external suggestions subject to social evaluation.

\subsubsection{AI as a Mirror}

For many participants, reading the AI reframes did not feel like receiving advice from AI, but rather as an extension of their own thought process. Participants articulated this experience through metaphors, describing the AI as a ``mirror'' reflecting their own thoughts and feelings back to them, with P19 noting: \textit{``Well, I guess it's like a mirror. It's mirroring my thoughts and feelings---but in a different way...''}. This conceptualization added to the therapeutic effect, making it easier for participants to internalize reframes. Additionally, in selectively reflecting thoughts, the AI mirror could validate and normalize experiences, and structure scattered ideas. All of these effects contributed to the therapeutic experience of AI-mediated reframing.

\paragraph{Fostering Internalization of Reframes}

First of all, the experience of reading reframed perspectives within their own first-person perspective often made it easier to internalize the reframes and apply them to their own situations. Participants described how reading the reframes using first-person pronouns made suggestions feel like they were \textit{``really coming from me''} (P3), \textit{``feel like a part of me''} (P24), \textit{``more intimate''} (P8), \textit{``more natural''} (P9). When asked to compare this with the experience of receiving reframes from a therapist, P23 explained that this experience both allowed them to express their thoughts more freely and made them feel more comfortable with the reframes:

\begin{quote}
``\textit{...I focused more on like just dumping my emotions and just putting it out there so like exactly like how you would use a journal. And then the fact that [the AI] continued with that tone, like with the first-person that I was using made it feel more natural... my writing style would be different if I was to interact with... a person, or a bot that was addressing what I was writing.}''
\end{quote}

This was particularly important for participants wary of social judgment. P9, an international student from a conservative country, contrasted the AI with his experience of judgmental mental health professionals, stating, \textit{``When it’s in the first person, it feels more like I’m hearing my own thoughts, and I don’t feel judged by them.''}.



\paragraph{Validating and Normalizing Experiences}

Although the primary aim of the reframing process was to generate new perspectives, participants also valued how the AI-reframed text incorporated their original situations, thoughts, and feelings. When participants saw their own thinking and experiences reflected and reiterated in the AI's output, they felt a sense of validation and normalization. This mirroring effect made participants feel understood and reassured that their feelings were legitimate. As P27 put it, \textit{``It kind of reinforces that point. It matches with my inner thoughts.''} Similarly, P19 described the comfort of seeing his thoughts and feelings echoed by the AI, linking this to the therapeutic benefit of feeling ``heard'', ``understood'' and ``normal'':

\begin{quote}
\textit{``It's mirroring my thoughts and feelings---but in a different way. So that's kind of encouraging in some way---oh, my thoughts are not stupid, they are solid. The AI can understand me, seems like I'm quite a normal person. It's validating my feelings, that it's normal to feel these feelings.''}
\end{quote}

The resonance was most effective when AI used nuanced language to describe these experiences, capturing the subtleties of an experience and demonstrating a deeper form of understanding. P8 was particularly struck by the AI's use of the word "often": \textit{``When it restates what I said, but in different words... I was amazed that some part of what I expressed was actually understood... That word---often---was emotionally accurate. If it had just said I don't, it would've felt too absolute... But `often' captures that gray area really well.''} For P8, this was the main therapeutic benefit: \textit{``I feel like I'm being heard. And then this thing is not stuck inside me anymore... that part of understanding already makes me feel like it's not that big of a deal.''}




\paragraph{Clarifying and Structuring Thoughts}

Participants described how AI-generated reframes clarified their experiences by helping them find the right words for complex or unarticulated emotions, as well as by summarizing and organizing their scattered thoughts, which could sharpen self-understanding. For instance, P16 describes difficulties she experiences when dealing with a past relationship. She had not consciously considered her relationship in terms of its benefits until the AI offered the word:

\begin{quote}
    \textit{``It gave me the word `\texttt{beneficial},' and that actually helped me reflect... Before reading this sentence, I had not thought about my relationship in that way.''} (P16)
\end{quote}

For others, the right words within the reframes helped them to ``\textit{reinforce the ideas I'm trying to hold onto right now}'' (P5) or made their own thoughts feel more concrete and better-phrased (P8). This was particularly helpful for participants like P23, who found it ``\textit{hard to name feelings}'' that were ``\textit{mesh[ed] together.}'' The AI could untangle these emotions, producing ``\textit{something distilled, pure}'' (P7) that captured what they ``\textit{wanted to express},'' and serving as ``\textit{reminders of how I've felt or what I've been going through emotionally}'' (P17).

This process often led to a deeper, more structured form of reflection, which was particularly beneficial when it alleviated the cognitive burden of articulating ideas. For example, P1 noted, ``\textit{When I'm writing, I feel I'm doing two demanding tasks at once: feeling, then figuring out how to phrase it. When I read the AI version, I can just reflect on the experience. It reduces mental load... Seeing the text reorganised lets me focus on the abstract reflection instead of the wording.}'' This sentiment was shared by others who found the AI's output to give them ``\textit{double chance to reflect... and see the overall picture —what the problems were, what caused those problems}'' (P3), or a helpful ``\textit{summary about what I'm thinking}'' (P24, P7), especially when their own thoughts felt rambling (P6).



\subsubsection{AI as Another Self}
\label{finding-ai-self}

Many participants described a sense of disconnection from the AI-generated reframed text, feeling that it no longer fully represented their own voice or identity. Instead, they experienced these AI reframes as if they were authored by a constructed, ``alternative'' self—an entity that drew upon their original words and, to some extent, their emotions, but ultimately felt separate and distinct from their authentic self:

\begin{quote}
    \textit{``Because it's all based on my writing, it's like seeing a very positive version of myself''} (P4) \\
    \textit{``It is like an aspired self. So it sort of makes it clearer that this is the person that I want to be.''} (P11)
\end{quote}
The experience of conceptualizing and interacting with this AI-generated self was profoundly double-edged. When it aligned with participants' desired qualities, it could be inspiring; but when it did not, it felt intrusive, inauthentic, and invalidating.

\paragraph{A Dialogue With an Inspired Self}

\begin{figure}[h]
    \centering
    \includegraphics[width=1\textwidth]{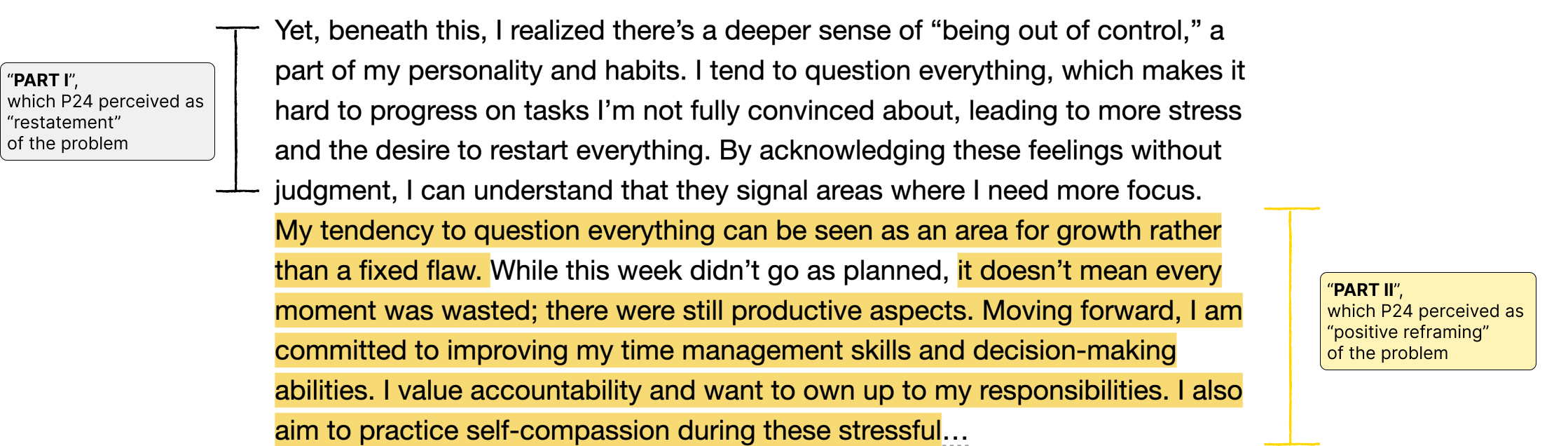}
    \caption{AI-reframing based on P24's narrative entry. The highlighted text maps to the texts that P24 chose to highlight during the think-aloud session. ``Part I'' refers to the sentences that P24 viewed as a ``restatement'' of the problem and ``Part II'' refers to the sentences that P24 perceived as ``positive reframing''.}
    \Description{One instance of an AI-reframed narrative that a participant (P24) perceived as having two distinct parts. Part I ("restatement"): The first paragraph of the text, which the participant saw as an acknowledgment and restatement of their original problem, including feelings of being "out of control."
Part II ("positive reframing"): The second, highlighted paragraph, which the participant saw as a constructive reframing. This section recasts the "tendency to question everything" as an "area for growth" and concludes with actionable goals and self-compassion.}
    \label{fig:P24-reframing}
\end{figure}

Some participants (P23, P24) experienced the reframes as a dialogue with a more positive version of themselves.
For example, P24 distinguished between the two parts of the reframed text: the AI's \textit{``restatement''} of her problem and the \textit{``positive reframing''} (see Figure \ref{fig:P24-reframing}). She described the latter part as the AI fostering a dialogue between her current self and a more aspirational one.

\begin{quote}
    ``\textit{The second part feels more like a conversation. I input some words, and it changes something for me. It’s not exactly my journal, but I do find it inspiring...because when I write, it’s mostly stating the facts...and then I’m done. But the AI extends that, and even though it’s written in the first person, it feels like... Like there’s an `inspired self', more positive self, giving feedback to me.}” (P24)
\end{quote}

This experience of the AI channeling an \textit{``inspired self''} made the reframed narrative feel personally resonant and motivating.

\paragraph{``Why are you trying to be me?''} 
\label{findings:trying-to-be-me}

When the AI's portrayal of ``another self'' felt misaligned, it created a sense of distance and intrusion. For example, P4 articulated: \textit{``On the other hand, it still feels like watching someone else. It's like watching myself act in a TV drama.''} The act of the AI \textit{``speaking as me''} was perceived as a violation of personal autonomy and emotional ownership. Participants were especially concerned when the AI, using a first-person voice, prescribed solutions or action plans—crossing a line from supportive reframing to unwelcome intrusion. P16 articulated this boundary:

\begin{quote}
\textit{``Personally, when the solution part is written using a lot of `I' sentences---as if it's speaking as me---then I start to feel like, `Why are you trying to be me?' Like, You don't have to give me tasks or tell me what I should do. I still want to have autonomy over how I feel and what I choose to do.''} (P16)
\end{quote}

Furthermore, the perception of the AI as ``another self'' became problematic was when it denied participant's feelings through perceived positivity. This created a sense of being misrepresented, as if the AI was ``\textit{speaking as me}'' (P16) or `\textit{twisting my words}'' (P15). Participants articulated that a narrative in their own voice creates an expectation of complete emotional alignment, whereas new perspectives would be more acceptable from an external (second-person) voice, like ``\textit{You’ve accomplished this much}'' and ``\textit{You've come this far}'' (P17). P8 further explained: \textit{``Right now, it feels like it’s putting words in my mouth. I want to have control over how my feelings are described. But I’m open to suggestions from another voice, as long as it doesn’t pretend to be me.''}.

Finally, feelings of intrusion were rooted in participants' awareness of the AI's inherent lack of the comprehensive, contextual knowledge required to authentically adopt their voice. This perceived pretense of understanding was felt as an emotional transgression, akin to being invalidated in a human conversation. For P13, the core problem was that the AI \textit{``knows nothing about me beyond the written text''}, while P8 framed it as an issue of interpersonal respect, explaining that the AI's first-person narration feels like a person presumptuously claiming total understanding.

\begin{quote}
    ``\textit{\texttt{`I have been grappling with a challenging decision.'} I'm not comfortable with that. [...] the biggest problem is that it addresses me in the first person. This large model knows nothing about me beyond the written text. }'' (P13) \\

    \textit{``If it's coming from another person, it might actually be easier to take in—because it doesn't feel like they're acting like they know everything I'm going through. That makes a big difference. I feel more respected emotionally. Sometimes, when people say, `Yeah, I know this,' I just want to respond, --- No—you don't really understand.''} (P8)
\end{quote}

\subsubsection{AI as a Social Actor}
\label{finding-ai-social}

Although AI incorporated the advice within the user's first-person perspective, many participants did not assign any "self" qualities to the reframes, and instead viewed it as AI's advice---a distinct external entity offering advice based on their original text.
\begin{quote}
    \textit{``Even when it’s written in first person, I don’t actually internalize it as me talking to myself. I still experience it as a separate entity---like someone else reflecting my thoughts back to me.''} (P18)
\end{quote}

Participants instinctively evaluated the AI’s output through the lens of familiar social dynamics and relational expectations, treating its contributions much like advice from a friend, a stranger, or even a psychologist. These attempts at positive reframing could often feel \textit{``patronizing''} through the lens of interpersonal norms:

\begin{quote}
\textit{``I think I can put [this reframe] as an example when you go to your friend, and you just wanna rant. And then they come to you, and they kind of give you advice or try to make you look at it in a positive way. Sometimes it makes you feel better. But sometimes it just feels patronizing.''} (P18)

\textit{``Recognize that my emotions are valid'—that’s like a psychologist telling me that. I already know that.''} (P26)
\end{quote}

AI's credibility and interactional boundaries were also shaped by the pervasive tendency to apply human interaction norms to AI-reframed text. For example, P15 described viewing the AI as a \textit{``stranger''}, which made its positive feedback feel less meaningful than feedback from trusted sources or from themselves. This perspective was reflected in P15's decreased positive affect and lower ratings of authenticity and ownership after receiving the AI's reframing. P15 shared:
\begin{quote}
    ``\textit{Positive reminders can work, but for me, it has to come from me. For example, if someone whose opinion I care about tells me I've been making good progress, I can take that to heart somewhat. But if a stranger says that, I don't actually care about what they think.}'' (P15)
\end{quote}

This evaluation was further rooted in fundamental beliefs about the AI's nature, as P15 concluded: ``\textit{If it's AI who says it, I really don't care about what it thinks, because it doesn't think.}''.

These instances demonstrate that perceiving the AI as an external social actor meant participants evaluated its input using social heuristics and assumptions about its authority and cognitive capacities (e.g. ability to understand), impacting their acceptance of its reframes.
\section{Discussion}
\label{discussion}

Our findings show that when AI intervenes in users' personal narratives for cognitive reframing, its role is not neutral. Users conceptualize the AI in three distinct ways---as a mirror, another self, or a social actor---which in turn shapes their experience of authenticity, narrative ownership, and the reframing process itself. Building on these findings, we first discuss the nuanced role of AI mediation within therapeutic narratives and its effects on the cognitive reframing process (Section \ref{discussion-mental-models}). Second, we outline how our findings can inform the design of future AIM-CR tools and broader implications for AI-based mental health interventions (Section \ref{discussion-designing}).

\subsection{Revisiting the Role of AI Mediation within AIM-CR}
\label{discussion-mental-models}

\subsubsection{The Spectrum of Perceived Self-Resemblance} Current research on AI-Mediated Cognitive Reframing (AIM-CR) often views the AI as a simple automation of a CR technique, reflected in its emotion-focused evaluations \cite{sharma2024facilitating, li2024skill}. Our work reveals a more complex psychological dynamic: users interact with AI reframes not simply as a tool, but develop distinct mental models of a ``mirror,'' an ``another self,'' or a ``social actor.''

We observed that users attributed these models according to the degree of self-resemblance---a perceived consistency between a user's sense of self and the AI-generated narrative. Here, by self-resemblance we mean the capacity of AI to offer constructive perspectives that aligned with users' values, and feel attainable (we distinguish this concept from sycophancy; a sycophantic AI would simply collude with a user's existing negative thoughts, which runs counter to the principles of CR\cite{10.1145/3715275.3732039}).

These findings resonate with existing research on ``self-similar media,'' where AI generates digital representations of the user like ``AI self-clones'' or ``digital doppelgangers'' \cite{10.1145/3579524, huang_mirror_2025}. Prior work highlights a key tension in these interactions. Engaging with one's digital likeness can foster a safe space for self-expression \cite{ta_user_2020}. On the other hand, when the AI generates statements users don't identify with, it can also provoke eeriness, threaten user autonomy, and create conflict between the desires for self-enhancement and authenticity \cite{10893443, huang_mirror_2025}. Below, we overview the unique dynamics specific to the therapeutic context of AIM-CR and provide speculations on their potential origins with references to psychological theories:

\begin{enumerate}
    \item When resemblance was \textit{high}, the AI reframes were similar to or aligned with users' existing perspectives, acting as a \textit{mirror}, which felt validating and affirming to the users. This dynamic can be understood through the lens of \textit{self-verification theory}, which posits that people seek to be known and understood in accordance with their self-views to maintain a coherent sense of self \cite{swann_jr_self-verification_2012}. 
    \item When resemblance was \textit{partial}, it became an aspirational but potentially alienating ``another self''---a version of the user that was more positive or capable. The therapeutic utility of this model was double-edged, and can be understood through Carl Rogers' theory of congruence, which links psychological well-being to the alignment between one's current and ideal self \cite{rogers1961becoming}. When the AI's ``aspired self'' felt attainable, it could bridge this gap, fostering congruence and motivating personal growth. However, when the projected self was too distant, it could widen the gap and created a state of incongruence.  
    \item When resemblance was \textit{low}, the AI was perceived as an external \textit{social actor}, whose advice was often scrutinized and dismissed. Once viewed as an AI (non-self), people effortlessly assign social characteristics to these tools---personalities, intentions and emotions \cite{krummheuer2015technical, nass1994computers}). This externalization led participants to judge the AI's credibility and intentions.
\end{enumerate}

\subsubsection{Beliefs about AI Technology Influence Engagement}

Participants' perceptions of the tool were heavily influenced by their prior understanding and expectations of AI. With the prevalence of AI systems like ChatGPT, users bring preconceived notions about AI's capabilities and personality to new applications, including self-help tools \cite{kobiella_if_2024, gessinger_chatgpt_2025, fang_how_2025}. For instance, the view of AI as non-judgmental and objective in our study aligns with broader societal perceptions of AI as objective and neutral \cite{10.1093/jcmc/zmz026}. However, negative beliefs also created barriers; one participant dismissed the AI, stating it ``\textit{does not think}'' and citing their knowledge of LLMs as next-word predictors.

Moreover, past experiences with AI raised expectations for the tool. Aware of AI's data-processing capabilities in tools like NotebookLM \footnote{https://notebooklm.google/} or ChatGPT with memory \cite{noauthor_memory_2024}, many users desired more personalization. Some would not trust the AI's advice without it having more context, such as access to past journals. Consequently, users' mental models of AI are a critical factor in their engagement with AIM-CR tools. Future designs should therefore incorporate these pre-existing beliefs and expectations, as we discuss in the next section.

\subsection{Designing AIM-CR Tools which Take AI Mediation into Account}
\label{discussion-designing}

Our findings reveal that users may spontaneously perceive AI's role differently within the text, which could affect their self-disclosure, authenticity, and other therapeutic outcomes. Designers of AIM-CR tools should therefore take these into account and consider how to 1) intentionally guide users towards appropriate mental models of the AI and its role 2) ensure users feel a sense of control over their narratives and reframes 3) where feasible and ethical, incorporate user context and pre-existing beliefs about AI.

\subsubsection{Guide Users Towards Appropriate Mental Models}

Each of these mental models offers unique therapeutic benefits, from validation and clarity to motivation and novel perspective-taking, but also introduce corresponding risks (see Section \ref{discussion-mental-models}). While these perceptions occurred spontaneously in our study, this presents a clear design opportunity: CR tool designers can intentionally tailor interventions to support the most appropriate mental model for a given therapeutic goal. Here, we overview how to approach this challenge.

\paragraph{Designing for the Mirror: Prioritizing Validation over Transformation} This ``mirroring''---reflecting user thoughts back to them in a validating, kind way --- was a powerful tool when the therapeutic goal is affirmation and emotional processing. This approach provided a safe, validating space while avoiding the risks of inauthenticity that arise when the AI's suggestions diverge from the user's present reality. To avoid risks of inauthenticity, designers could focus reframing techniques on acknowledgment, validation, and structured reflection of the user's narrative (e.g. from Positive Psychology \cite{seligman2000positive}), rather than introducing new, transformative perspectives.

\paragraph{Designing for the Aspired Self: Co-Creating Motivation with Agency} Building another self model---introducing novel, actionable, aspired qualities---was a powerful motivator, helping users envision a "possible self" that could spur them to action. However, it introduced tension and feelings of intrusion due to a perceived lack of agency in shaping this aspirational narrative. The AI appeared to be ``pre-committing'' the user to a future self they had not consented to. Therefore, when the goal is motivation and growth, the system must allow the user to co-create the Aspired Self. Rather than presenting a finished reframe, the system should give users explicit, granular control over the transformation to ensure the aspired self aligns with the user's preferences. This could be implemented by: (1) allowing users to choose from multiple reframes with varying degrees of optimism; (2) providing controls to adjust the tone or specific wording of suggestions; or (3) prompting users to define their core values, which then guide the AI's reframing process. For example, In Pataranutaporn et al.'s research, to reduce gaps between current self and the alternative self---a key tension in our study---users could specify diverse additional qualities, like proud memories, turning points, and future lifestyles \cite{10893443}. This transforms the interaction from a passive reception of an AI's ideal to an active collaboration with a tool for self-development.

\paragraph{Designing for the Social Actor: Building Trust for Different Perspectives} When the AI's suggestions were highly novel or distant from the user's current self-concept, our participants suggested that having the AI adopt an external, ``social actor'' role (addressing the user through a second-person pronoun) is a preferred mode of interaction. However, it is important to build trust. ``Social actor" was often perceived negatively in our study because it was an untrustworthy stranger, its advice felt generic or ``patronizing''. Therefore, when reframes introduce very different perspectives or advice, the system must be designed as a trustworthy conversational partner. The system should make AI's presence explicit with proper cues (e.g. personification, seamful interaction \cite{chalmers_seamful_2003}) and build credibility. For example, this credibility can be achieved by transparency and explainability \cite{heinz2025randomized}---the AI may explain its therapeutic rationale for deriving the reframes through evidence-based psychological frameworks, for example, Cognitive Behavioral Therapy.




\subsubsection{Ensure User Agency in Narrative Co-construction}

In cognitive reframing, new perspectives have to be integrated into existing narratives about the self \cite{beck2011cbt}---they have to resonate with the person's lived experiences and internal sense of truth. A disconnect between the two can create ``authenticity gaps'', a distance between a plausible reframe and one that is deeply felt and personally meaningful \cite{anderson2020does}. During our study, participants actively edited the outputs to align it with their emotional reality and personal voice. For example, to emotionally align the AI's reframes, participants softened overly assertive language, corrected misinterpretations, and preserved essential emotional complexity, such as the ``jagged parts'' of their story. They also differed in their preferences for the reframing style---some preferred validation-focused reframes, while some preferred actionable advice. This is in contrast to many AIM-CR works, which use predetermined psychological frameworks (CBT, positive psychology) to guide the reframing process \cite{ziems2022inducing, sharma_cognitive_2023, zhan2024large, li2024skill}. 



Therefore, one design challenge for AI-assisted reframing tools is allowing for user agency in modifying and adjusting the AI reframes to their self-concepts, current emotional states, and personal goals within the existing reframing frameworks. In this way, the would AI catalyze the users' reflective work rather than fully automating the process. We consider how future AIM-CR tools can support user agency and personalization:




\paragraph{Allowing the Choice of Reframing Styles} Participants varied in the kind of support they sought—some appreciated optimistic reinterpretations, others preferred practical actions, or even reframes that \textit{preserved} the intensity of their original emotions. These differences were often related to current mood and personal coping style, consistent with findings that individual traits and situational factors jointly shape reappraisal outcomes \cite{troy2013person}. To accommodate this variability, systems might allow users to select or declare their preferred reframing approach (for example, cognitive-behavioral or acceptance-based strategies). Some mental health agents can be explicitly adapted to personality \cite{ahmad2022designing}.

\paragraph{Allow for Granular User Choices and Collaborative Reframing} Participants in our study used their agency to actively edit the AI's outputs, softening assertive language, correcting factual misinterpretations, and altering the tone to fit their personal style. Given these diverse preferences, systems should offer explicit controls to accept, reject, or modify suggestions, fitting them to a user's personal context.

\paragraph{Personalize the Tone} For reframes to feel authentic, they must align with a user's personal style—a quality that encompasses word choice but also emotional nuance and narrative complexity. While basic prompting can improve stylistic alignment \cite{bhandarkar2024emulating}, deeper personalization can be achieved by implicitly modeling a user's tone from their texts \cite{yeh2024ghostwriter}. This method has been shown to enhance stylistic resonance and foster a stronger sense of ownership over the generated text \cite{yeh2024ghostwriter}.

\subsubsection{Incorporate User Context}

Finally, many participants stressed that effective reframes should not remain ``\textit{generic}'', but rather directly relate to their lives by acknowledging their specific experiences, patterns, and challenges. This aligns with prior research on cognitive reappraisal, which emphasizes that a strategy's success depends heavily on its fit with both the person and the situation: what soothes or motivates in one circumstance may feel unhelpful, or even aggravating, for another \cite{troy2013person, sheppes2014emotion}. Recent research offers examples of how user's experiences and contexts can be integrated into AI support. For example, \textsc{MindScape} used recent smartphone-sensed behaviors to ground journaling reflection prompts in the user's current lived context \cite{nepal2024mindscape}; \textsc{Resonance} used the user's past journals to craft future-focused suggestions that echo earlier entries, improving affect and engagement \cite{zulfikar2025resonance}. Given that CR is often a long-term process, it may be possible to reuse previous entries and situations to generate new reframes, an approach similar to what was done in \textsc{DiaryMate} \cite{kim2024diarymate}.

\subsection{Limitations}

While this study offers valuable insights, its limitations highlight avenues for future research. Our findings' generalizability is constrained by our sample of 28 participants, 27 of whom were graduate students, from the northeastern U.S. without clinical disorders. Their experiences may not reflect those of broader demographic, cultural, or clinical populations, particularly individuals facing severe psychological distress. Future work should evaluate AI-mediated Cognitive Reframing (AIM-CR) in more diverse and clinical settings.


Additionally, the single-session design prevents conclusions about AIM-CR's long-term effects, such as sustained engagement or evolving user perceptions. Longitudinal studies are essential to track how user interactions, trust, and perceived utility evolve with continued use. Finally, our results were also shaped by the specific AI model (GPT-4o) and minimalist interface used; different technologies or more interactive designs could yield different outcomes. Longitudinal and comparative studies are needed to explore these variables and guide the design of more effective, ethical, and user-centered tools.


\section{Conclusion}

This study demonstrates that AI-mediated Cognitive Reframing (AIM-CR) significantly reshapes the intrapersonal process of self-reflection, revealing that AI mediation is not a neutral automation but a transformative intervention that deeply affects user experience. Users spontaneously conceptualized the AI as a \textit{Mirror}, an \textit{Aspired Self}, or a \textit{Social Actor}, based on the perceived alignment between AI-generated suggestions and their self-concept. Each model offered distinct therapeutic benefits and risks: \textit{Mirror} provided validating self-verification; \textit{Aspired Self} presented motivating visions that could evoke inadequacy; and \textit{Social Actor} introduced novel but potentially untrustworthy perspectives. Crucially, the AI's ``voice'' emerged as pivotal in shaping authenticity, narrative ownership, and the therapeutic alliance. Effective AIM-CR thus requires nuanced design that intentionally steers users toward beneficial mental models, grants meaningful control for narrative co-construction, and builds trust through contextual sensitivity. This highlights that therapeutic AI depends not only on reframing text but sensitively navigating the delicate human-AI relationship.

\begin{acks}

We would like to acknowledge the use of AI language models, including ChatGPT and Google Gemini, to assist with proofreading, language polishing, and formatting of certain sections of this paper. The AI tools were used to improve readability and presentation while all substantive content, research design, data analysis, and conclusions remained the exclusive intellectual contribution of the human authors, in accordance with ACM's Policy on Authorship.


\end{acks}

\bibliographystyle{ACM-Reference-Format}
\bibliography{references_z, references_o}

\clearpage
\appendix

\section{Writing Prompts for Users}
\label{appendix:writing-prompt}

\begin{figure}[h]
    \centering
    \includegraphics[width=0.75\textwidth]{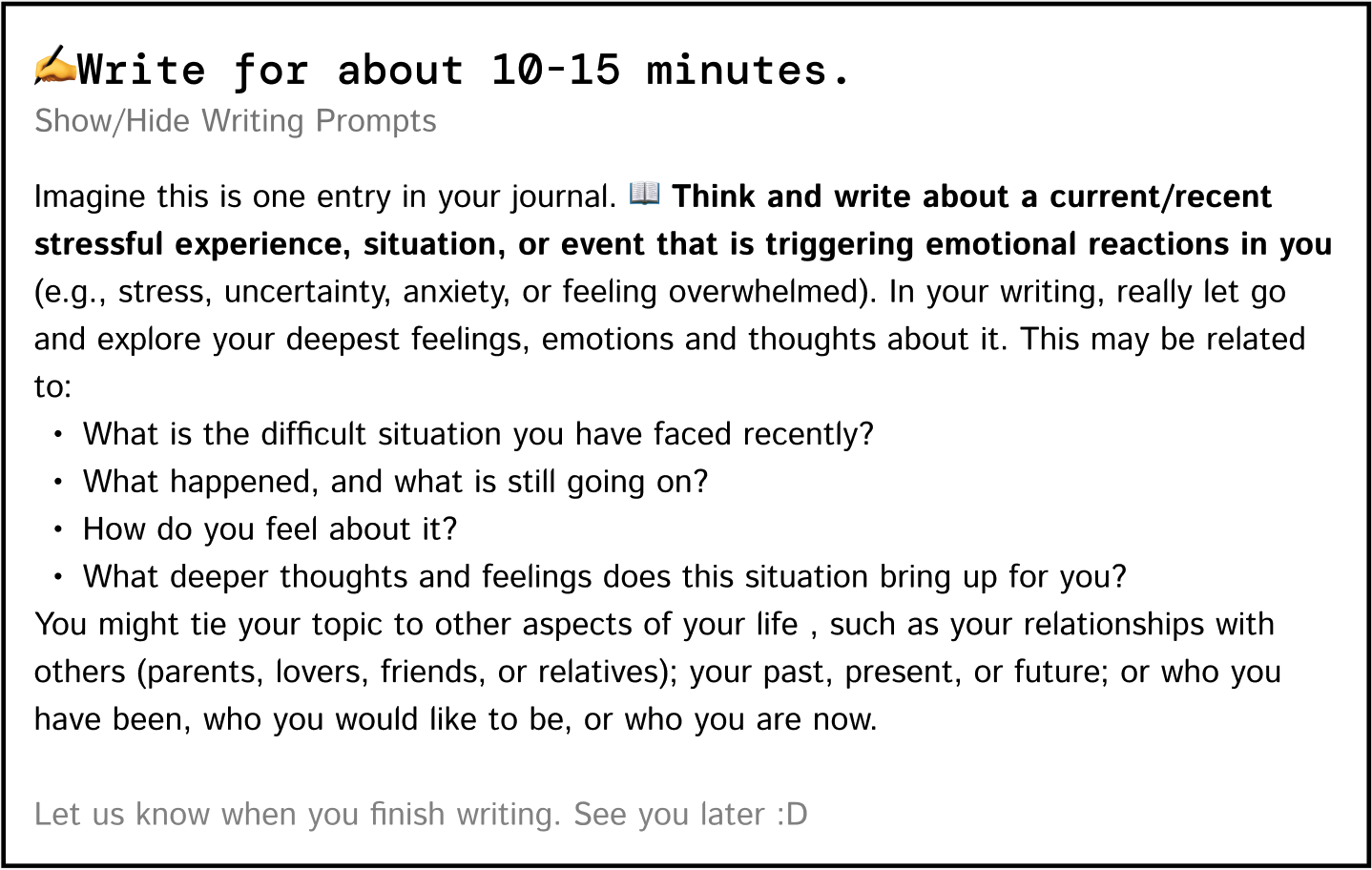}
    \caption{Writing prompt that was shown to the participant during the study. The writing prompt is adapted from the Expressive Writing framework \cite{pennebaker_expressive_2014}. }
    \Description{}
    \label{fig:writing-prompt}
\end{figure}



\section{AI Prompts}
\label{appendix:ai-prompt}
The main content (without redundant delimiters) of the original prompt used in our technology probe is shown below in Table \ref{table:ai-prompt}.

\begin{table}[h]
\centering
\begin{tabular}{|p{0.5\linewidth}|p{0.45\linewidth}|}
\hline
\textbf{Prompt by Intermediate Steps} & \textbf{Few-Shot Example} \\
\hline

\cellcolor{yellow!20} \textbf{System}: I have a personal journal that I want to rewrite using principles of evidence-based therapy to make it more motivational. When you rewrite, pay attention to the author's original style and tone of writing and preserve it naturally. & [\textit{input}] Original Thought: \{\textbf{I just don’t feel like the meaning of the research I'm doing.}\}\\ 
\hline
Identify and list the key thoughts and feelings in the text. & \draft{N/A} \\
\hline
   Identify any cognitive distortions (e.g., all-or-nothing thinking, catastrophizing, overgeneralization, personalization, black-and-white thinking, should statements). & Identify Cognitive Distortion: This might be a form of all-or-nothing thinking or catastrophizing.\\
\hline
Challenge the thoughts: Ask questions to verify the accuracy of the thoughts and provide alternative perspectives. &   Challenge the Thought: Is it true that your research has no meaning, or is it that the current situation feels particularly challenging? \\
\hline
Reframe the thoughts: Replace negative or distorted thoughts with more balanced and constructive ones. &  Reframe the Thought: While I'm currently feeling disconnected from my research, this may be a temporary phase. I can find ways to infuse meaning into my work or explore other opportunities that align better with my values.
\\
\hline
 Acknowledge and accept thoughts and feelings. &  Acknowledgement: the feeling of disconnection. \\
\hline

Identify several main values underlying the person's thoughts and feelings, and suggest ways to act in accordance with those values. & Identify what you value in research (e.g., creativity, autonomy). Goals  or ways to act corresponding to the values: seek ways to integrate these values into your current work or future projects.\\

\hline

\cellcolor{yellow!20} Rewrite the text by integrating the revised thoughts and feelings from previous steps. Ensure the rewritten text is more positive, motivating, and retains the essence of the original content. &  [\textit{output}] \{\textbf{I can seek meaningful experiences and connections in my daily life while pursuing my professional goals.}\}\\

\hline

\end{tabular}
\caption{Prompts were used with the GPT-4o model to generate reframing. The output was formatted on the backend to present only the final reframed narratives.}
\label{table:ai-prompt}
\end{table}




\section{Pre-Intervention and Post-Intervention Measures}
\label{appendix-measures}

To gauge how the intervention affected the participants, we used a series of pre-intervention and post-intervention measures. These included:

\textbf{Emotional Intensity} asked the participant which emotion the situation they described in the writing elicited, and they were asked to gauge the intensity of the emotion (scale from 0--10).

The \textbf{Positive and Negative Affect Schedule (PANAS)} encompasses two subscales to assess positive and negative emotions, asked as 5-item questions (scale 1--5). Participants indicated the extent to which they felt each given emotion (e.g., "Ashamed" or "Upset").






\textbf{Authenticity} included two questions (scale 1--5) pertaining to current-self authenticity ("I believe this text represents my current self") and aspired-self authenticity ("I believe this text represents my aspired self").


\textbf{Ownership} evaluates the extent to which participants identify the text as their own writing (scale 1--5). The question asked: "I identify this text as my own writing."










\end{document}